\documentclass[aps,pre,superscriptaddress,showpacs,floatfix,twocolumn,amsmath,amssymb]{revtex4}

\usepackage[]{graphicx} 
\usepackage{subfigure}
\usepackage{amsmath}
\usepackage{amssymb}
\usepackage{epsfig}
\usepackage{psfrag}

\def\be{\begin{equation}}
\def\ee{\end{equation}}
\def\bea{\begin{eqnarray}}
\def\eea{\end{eqnarray}}
\def\II{\hbox{$1\hskip -1.2pt\vrule depth 0pt height 1.6ex width 0.7pt\vrule depth 0pt height 0.3pt width 0.12em$}}

\newcommand{\refsec}[1]{\mbox{Sec.~\ref{#1}}}
\newcommand{\reftab}[1]{\mbox{Tab.~\ref{#1}}}

\begin{document}

\hyphenation{re-so-nan-ce re-so-nan-ces ex-ci-ta-tion z-ex-ci-ta-tion di-elec-tric ap-pro-xi-ma-tion ra-dia-tion Me-cha-nics quan-tum pro-posed Con-cepts pro-duct Reh-feld ob-ser-va-ble Se-ve-ral rea-so-nable Ap-pa-rent-ly re-pe-ti-tions re-la-tive quan-tum su-per-con-duc-ting ap-pro-xi-mate cri-ti-cal func-tion wave-guide wave-guides}

\title{Spectral Properties and Dynamical Tunneling in Constant-Width Billiards}
\author{B. Dietz}
\email{dietz@ikp.tu-darmstadt.de}
\affiliation{Institut f\"ur Kernphysik, Technische Universit\"at Darmstadt, D-64289 Darmstadt, Germany}
\author{T. Guhr}
\affiliation{Fakult\"at f\"ur Physik, Universit\"at Duisburg-Essen, Lotharstra\ss{}e 1, D-47048 Duisburg, Germany}
\author{B. Gutkin}
\affiliation{Fakult\"at f\"ur Physik, Universit\"at Duisburg-Essen, Lotharstra\ss{}e 1, D-47048 Duisburg, Germany}
\author{M. Miski-Oglu}
\affiliation{Institut f\"ur Kernphysik, Technische Universit\"at Darmstadt, D-64289 Darmstadt, Germany}
\author{A. Richter}
\affiliation{Institut f\"ur Kernphysik, Technische Universit\"at Darmstadt, D-64289 Darmstadt, Germany}

\begin{abstract}
We determine with unprecedented accuracy the lowest $900$ eigenvalues of two quantum constant-width billiards from resonance spectra measured with flat, superconducting microwave resonators. While the classical dynamics of the constant-width billiards is unidirectional, a change of the direction of motion is possible in the corresponding quantum system via dynamical tunneling. This becomes manifest in a splitting of the vast majority of resonances into doublets of nearly degenerate ones. The fluctuation properties of the two respective spectra are demonstrated to coincide with those of a random-matrix model for systems with violated time-reversal invariance and a mixed dynamics. Furthermore, we investigate tunneling in terms of the splittings of the doublet partners. On the basis of the random-matrix model we derive an analytical expression for the splitting distribution which is generally applicable to systems exhibiting dynamical tunneling between two regions with (predominantly) chaotic dynamics. 
\end{abstract}
\pacs{03.65.Sq, 05.45.Mt, 41.20.Jb, 73.40.Gk, 74.50.tr,}
\maketitle

\section{\label{Intro}Introduction}
The spectral fluctuation properties of a generic  quantum system with chaotic classical dynamics are known to be universal and to depend only on the associated symmetry class \cite{Casati1980,Berry1981a,Bohigas1984}. Accordingly, those of the eigenenergies of spinless, chaotic systems possessing time-reversal invariance are well described by the Gaussian Orthogonal Ensemble (GOE) of random matrices~\cite{Mehta1990,Guhr1998}. On the other hand, if time-reversal invariance is absent, the eigenvalue statistics coincides with that of random matrices from the Gaussian Unitary Ensemble (GUE). Indeed, such a spectral behavior  is  observed in  generic chaotic
systems, but not necessarily if additional discrete  symmetries are present.
A well known example is provided by  time-reversal invariant systems with rotational symmetries. For instance, the  spectrum of  billiards with  threefold symmetry can be split into singlets and doubly degenerate levels. While the former follow GOE statistics, the latter, in fact, behave like the eigenvalues of random matrices from the GUE~\cite{Leyvraz1996,Keating1997,Dembowski2000}.

A different dynamical mechanism for  the ``breaking'' of the time-reversal invariance was proposed in Ref.~\cite{Gutkin2007}, where quantum billiards defined on a domain of constant width with smooth boundaries were considered. By virtue of the special geometry of these billiards their classical phase space is split into two ergodic components: a particle launched into the billiard in clockwise and anticlockwise direction, respectively, moves in that rotational direction forever~\cite{Knill1998}. They correspond to two parts of the phase space that are well seperated by a region of Kolmagorov-Arnold-Moser (KAM) tori and possess an almost fully chaotic dynamics. Other examples of constant-width billiards where two ergodic components with fully chaotic dynamics  are separated in phase space by just one singular line are given in Refs.~\cite{Veble2007,Gutkin2009, Bunimovich2009}.

Such a remarkable  separation  of the phase space, in turn, leads to extraordinary properties of the corresponding quantum billiard. In particular, its spectrum can be split into quasi-degenerate doublets and singlets. Furthermore, for billiards of constant width with smooth boundaries the spectral statistics of the doublets was shown to be of GUE type rather than of GOE type~\cite{Gutkin2007}. This indeed is reminiscent of the spectral structure in systems with threefold symmetries. However, contrary to the latter, the energy levels in the doublets of constant-width billiards are not fully degenerate but rather have very small splittings as compared to the mean spacing of the doublets. While the switching between clockwise and  anticlockwise motion is classically strictly forbidden, it is possible via tunneling through the barrier of KAM tori in the quantum billiards. This effect, known as dynamical tunneling, manifests itself in the eigenenergy splittings within the doublets.

Dynamical tunneling has been extensively investigated in the last two decades. Reference~\cite{Keshavamurthy2011} provides a detailed survey on current theoretical and experimental studies within that field of research. So far, most of the works have focussed on tunneling of quantum particles from regular to chaotic regions of phase space. To describe this phenomenon, various methods based both on semiclassical approaches and on Random Matrix Theory (RMT) have been developed~\cite{Baecker2008,Backer2010,Loeck2010,Mertig2013}. The regular-to-chaotic tunneling process is also a building block of the  theory of chaos-assisted tunneling~\cite{Bohigas1993,Wilkinson1996,Leyvraz1996a,Schlagheck2006,Mouchet2001,Mouchet2003} which describes tunneling between two regions with regular dynamics through one with chaotic dynamics. First experimental evidence for chaos-assisted tunneling in a flat microwave annular billiard has been presented in Refs.~\cite{Hofferbert2000, Hofferbert2005}. For the billiards of constant width the situation is in fact opposite. Here, a quantum particle tunnels through a dynamical barrier of regular KAM tori between two regions with chaotic dynamics. To the best of our knowledge such a chaos-to-chaos scenario has been studied so far only for potential barriers~\cite{Creagh1996}.

Since the tunneling is reflected in the size of the energy splittings, an understanding of their statistical properties is of great interest. The main goal of the present work is to obtain the energy levels of two billiards of constant width experimentally and to study the distribution of the doublet splittings. It should be noted that in the semiclassical limit the splittings become very small in comparison to the mean spacing of the doublets. Accordingly, the experimental determination of the eigenvalues is very challenging because it requires very precise measurements which we achieved by using superconducting microwave billiards. We present a RMT model for the splitting distribution that, as is demonstrated in the present article, describes the experimental results very well.

The article is organized as follows. In~\refsec{Const} we summarize the salient properties of the classical dynamics of constant-width billiards. Section~\ref{Exp} comprises the description of the microwave experiments. We used the resonance frequencies extracted from the experimental spectra to compute the associated wave functions and the Husimi functions. This is outlined in~\refsec{Num}. In ~\refsec{Fluct} we show results for the spectral properties of two microwave constant-widths billiards and in~\refsec{Stat} for their splitting distributions. The RMT model developed to describe the latter is introduced in~\refsec{RMT}, and the analytical results derived from it in ~\refsec{Analytic}. Finally, we present our conclusions in~\refsec{Concl}.     
\section{\label{Const}Constant-Width Billiards}
\begin{figure}[tb]
    \includegraphics[width = 0.5\linewidth]{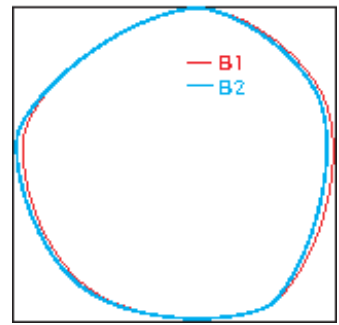}
    \caption{(Color online) Sketches of the two billiard shapes. That of billiard B1 is shown in red (dark gray), that of B2 in blue (light gray). They are barely distinguishable.}
      \label{fig:shapes}
\end{figure}

\noindent
The billiards under consideration are convex and have a constant width, i.e., such a billiard can be inserted in a square which touches it at each of its side, and this still will be the case when the billiard is rotated. The points $(x(\alpha),y(\alpha))$ on the boundary can be expressed as the real and imaginary parts of the complex function~\cite{Knill1998}
\begin{equation}
z(\alpha)=-ia_0-i\sum_{n\in\mathbb{Z}}\frac{a_n}{n+1}\left(e^{i(n+1)\alpha}-1\right),\ \alpha\in [0,2\pi)
\label{boundary}
\end{equation}
where $a_{-n}=a^*_n$, $a_1=0$ and $a_{2n}=0$ for $n>0$. The width, i.e., the diameter and the perimeter equal $2{\rm R}=2a_0$ and $2\pi a_0$, respectively. We investigated the properties of two constant-width billiards, called B1 and B2 in the following. The sets of coefficients $a_n$ were chosen as listed in~\reftab{table0}. 
\begin{table}[t]
\caption{List of the coefficients $a_n$ defining the boundaries Eq.~(\ref{boundary}) of billiards B1 and B2.}
\label{table0}
\begin{tabular}{c|c|c|c|c|c|c}
\hline
\hline
   & $a_0$ & $a_1$ & $a_3$ & $a_5$ & $a_{2n},\, n\geq 1$ & $a_{2n+1},\, n\geq 3$\\
\hline
B1 & $12$ & $0$ & $\frac{3}{2}i$ & $3$ & $0$ & $0$ \\
B2 & $12$ & $0$ & $2i$ & $3.9$ & $0$ & $0$ \\
\hline
\hline
\end{tabular}
\end{table}
A sketch of the billiard shapes is shown in Fig.~\ref{fig:shapes}.

\begin{figure}[tb]
        \centering
        \includegraphics[width = \linewidth]{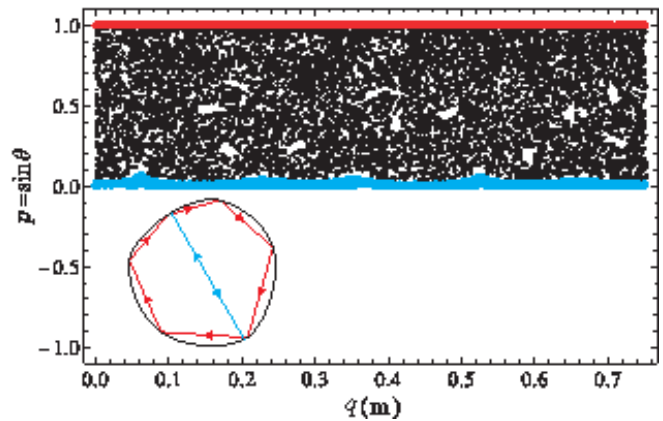}
\caption{(Color online) PSOS for billiard B1. For its construction particles were launched into the billiard in clockwise direction. Due to unidirectionality only the upper half ($p>0$) of the PSOS is filled. The chaotic sea is bordered by a region of whispering gallery orbits around $p=1$ plotted in red (dark gray) and a region of regular orbits around the BBOs at $p=0$ depicted in blue (light gray). A zoom into these regions is shown in Fig.~\ref{fig:whisp} for B1 and B2. The inset shows an example for a whispering gallery orbit in red (dark gray) and a BBO in blue (light gray).} 
        \label{fig:psos}
        \end{figure}

Constant-width billiards have the particular property that a particle launched into the billiard in a certain rotational direction will always rotate in the same direction, that is, the classical dynamics is unidirectional. This becomes obvious when looking at the Poincar{\'e} surface of section (PSOS), shown in Fig.~\ref{fig:psos}. Plotted is the sinus of the angle between the particle trajectory and the normal to the boundary at the point of impact, $p=\sin\theta$, versus the location $q$ of the latter measured in units of the arclength along the boundary. 
For the construction of the PSOS all particles were launched into the billiard in clockwise direction, so due to the unidirectionality only the upper half of the PSOS ($p>0$) is filled. As is visible in Fig.~\ref{fig:psos}, the PSOS of B1 consists of a large chaotic sea (black dots) but also contains two regions of regular KAM tori: one with whispering gallery orbits close to $|p|=1$ shown in red (dark gray), and the barrier region around the diameter orbits at $p=0$ depicted in blue (light gray). We call the latter in the following ``Bouncing Ball Orbits'' (BBOs) because they bounce back and forth between two opposite sides of the billiard~\cite{Shudo1990}. Thus the classical dynamics is mixed regular / chaotic. Examples for a whispering gallery orbit and a BBO are depicted in the inset of Fig.~\ref{fig:psos}.

\begin{figure}[tb]
        \centering
        \includegraphics[width = \linewidth]{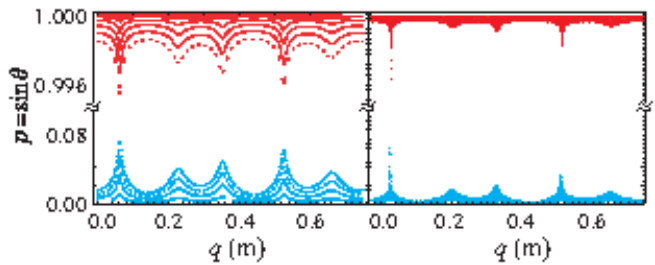}
\caption{(Color online) Regular regions in the PSOS of B1 (left) and B2 (right). Shown are a zoom into the region around the whispering gallery orbits ($p\simeq 1$) and one into that around the BBOs ($p\simeq 0$).}
        \label{fig:whisp}
        \end{figure}

\begin{figure}[tb]
        \centering
        \includegraphics[width = \linewidth]{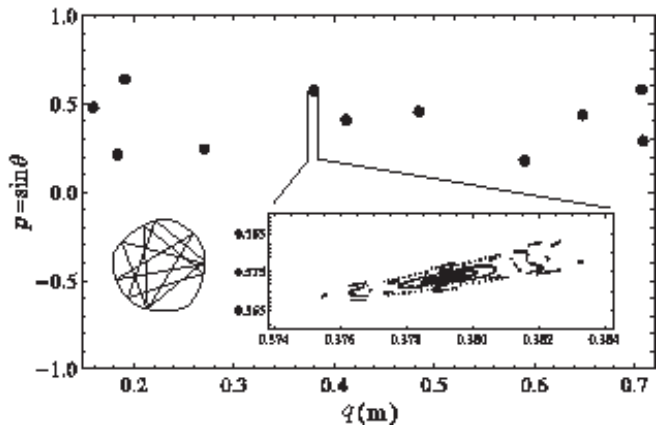}
\caption{Small regular islands in the chaotic sea of the PSOS (black spots). The left inset illustrates the periodic orbit corresponding to them. Only this one was found. For the billiard used in the experiment it has a length of $2.305$~m. The right inset shows a zoom into one of the small islands.}
        \label{fig:insel}
        \end{figure}

Although the shapes of B1 and B2 are barely distinguishable, their PSOSs differ in the widths of the regular regions, that are larger in B1 as demonstrated in Fig.~\ref{fig:whisp}. 
For that billiard we in addition found very small regular islands in the chaotic sea that are marked by black spots in Fig.~\ref{fig:insel}. They correspond to one regular orbit with 11 reflections and a length of $2.305$~m for the billiard geometry used in the experiment. It is shown together with a zoom into one such island in the left and the right inset of the figure, respectively. We only found this regular orbit with islands in the chaotic sea for B1, and none for B2. Thus, if there are any at all for B2, then they are even smaller than for B1. 

\section{\label{Exp}Experiment}
The change from clockwise to anticlockwise motion, that is, the transition from positive to negative momentum and vice versa is classically forbidden. In the quantum billiard, however, this is possible via dynamical tunneling~\cite{Davis1981} through the KAM tori barrier in the vicinity of the line $p=0$. The aim of the present work was the investigation of the spectral properties and the dynamical tunneling of quantum billiards with the shapes of B1 and B2. 

We determined the eigenvalues experimentally using flat cylindrical microwave resonators. Below a certain excitation frequency of the microwaves sent into such a resonator, which corresponds to a wave length equal to twice the height of the resonator, the electric field strength is parallel to the cylinder axis. Then the underlying Helmholtz equation is mathematically equivalent to the Schr\"odinger equation of the quantum billiard of corresponding shape with Dirichlet boundary conditions~\cite{Stoeckmann1990,Sridhar1991,Graf1992}. Accordingly the eigenvalues and the eigenfunctions of the latter are directly related to the resonance frequencies and the electric field strengths inside the microwave resonator, respectively, which is therefore called microwave billiard. 

\begin{figure}[tb]
        \centering
        \includegraphics[width = 0.5\linewidth]{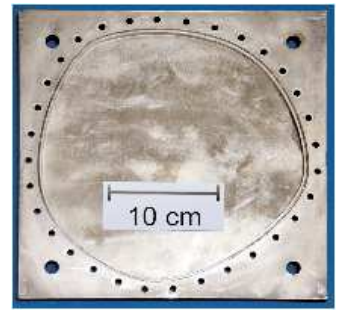}
\caption{(Color online) Photo of the microwave cavity with the shape of B1. It was constructed from copper plates. A hole with the shape of the billiard was milled out of the bottom plate. The top plate has been removed. For the measurements both plates were lead-plated and screwed together tightly through the holes along the cavity boundary.}
        \label{fig:photo}
        \end{figure}

The microwave billiards were constructed from two copper plates, where a hole with the shape of the respective billiard was milled out of the bottom one. A photo of the cavity with the shape of B1 is shown in Fig.~\ref{fig:photo}.
For both billiards the diameter was chosen equal to $24$~cm. The depth of the holes and thus the height of the resonators was 5~mm which corresponds to a maximum frequency of 30~GHz. 

The eigenvalues were determined from the positions of the resonances in the transmission spectra that were measured with high precision at superconducting conditions. In order to attain superconductivity at liquid helium temperature both plates were lead-plated and screwed tightly together through holes along the boundary (see Fig.~\ref{fig:photo}), thus achieving a quality factor $Q\approx 10^6-10^7$. For the measurement of the resonance spectra two antennas were attached to the top plates and microwave power was coupled into the resonators through one antenna and coupled out either via the same or via the other one. A vector network analyzer measured the relative phase and amplitude of the output to the input signal. This yields the scattering matrix elements $S_{11},\, S_{12},\, S_{21}$ and $S_{22}$ describing the scattering processes with the antennas $1$ and $2$ acting as single scattering channels. 

\begin{figure}[tb]
        \centering
        \includegraphics[width = \linewidth]{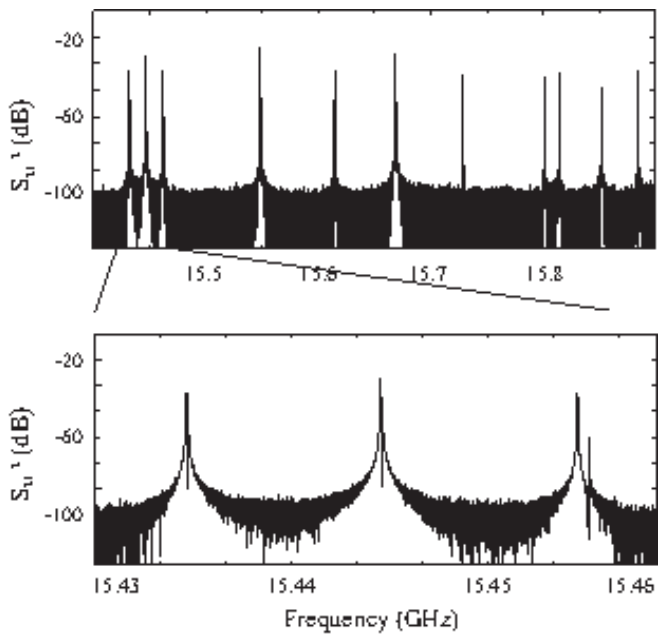}
\caption{Resonance spectrum measured with B1. The upper panel shows it in a frequency range from $15.4-15.9$~GHz, the lower one a zoom into a triplet of nearly degenerate pairs of resonances.}
        \label{fig:spectra}
        \end{figure}

We determined the resonance frequencies up to 25~GHz. Part of the spectrum measured with B1 is shown in the upper panel of Fig.~\ref{fig:spectra}.
At first sight the spectrum seemed to consist of single, sharp resonances. However a closer look revealed that almost all resonances are split into doublets (see lower panel of Fig.~\ref{fig:spectra}). This quasidegenerate structure is attributed to the tunneling from one rotational direction to the other one. Here, one doublet partner corresponds to the symmetric, the other one to the antisymmetric combination of clockwise and anticlockwise quasimodes that have  Husimi distributions localized in the upper and the lower part of the PSOS~\cite{Gutkin2007}, respectively. In addition, we observed a few singlets that are associated with the BBOs located along the $p=0$ line in the PSOS. For both billiards the resulting eigenvalue spectra consisted of $15$ singlets and $\approx 390$ doublets.

\begin{figure}[tb]
        \centering
        \includegraphics[width=\linewidth]{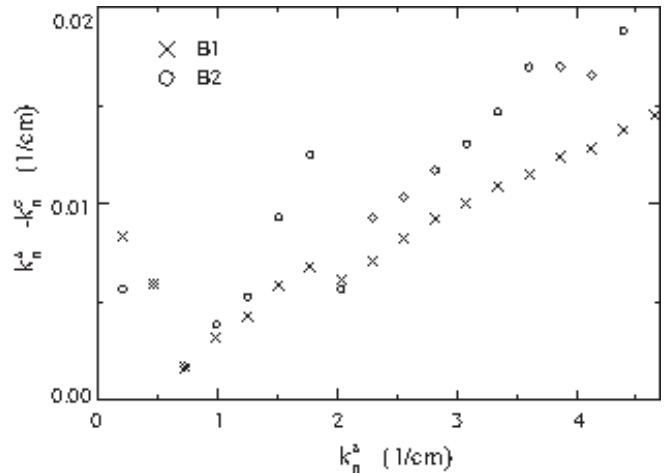}
        \caption{Differences between the zeroes of the $J_0$ Bessel function and the measured resonance frequencies of the singlets for B1 (crosses) and B2 (circles). Shown are the absolute values of the differences of the corresponding wave numbers $k$, where $k=2\pi f/c$.}
        \label{fig:j0}
\end{figure}

The singlets are analogous to the zero-momentum modes of the circle quantum billiard with the same diameter $2{\rm R}=2a_0$~\cite{Berry1981}. The eigenfunctions of the latter correspond to $J_0(k^c_nr)$ Bessel functions with the eigenvalues $k^c_n$ determined by the Dirichlet boundary condition $J_0(k^c_n{\rm R})=0$. Figure~\ref{fig:j0} shows the deviations of the values of the wave numbers $k^s_n=2\pi f^s_n/c$, with $f^s_n$ denoting the resonance frequencies of the singlets in B1 and B2, from the wave numbers $k^c_n$. 
They are slightly larger for B2. This is expected, because in its PSOS the width of the regular region around the BBOs is smaller than for B1, indicating that its shape corresponds to a more deformed circle.

\section{\label{Num}Numerical computation of the wave functions and Husimi functions}
\begin{figure}[tb]
        \centering
        \includegraphics[width=\linewidth]{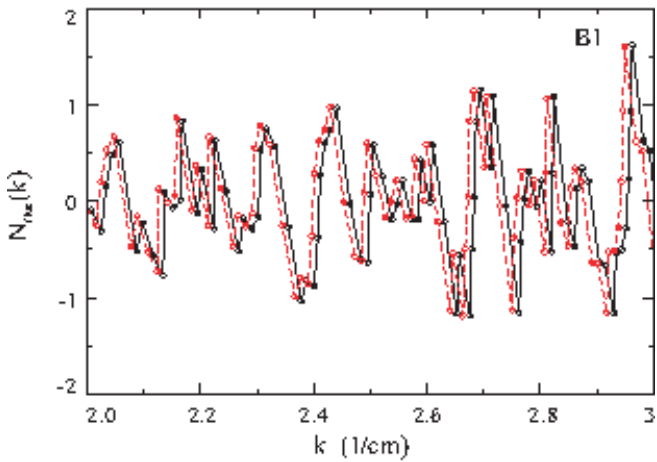}
\caption{(Color online) Comparison of the experimentally (black full line) and the numerically (red dashed line) determined fluctuating parts of the integrated resonance density for B1. This was done to identify in the list of possible eigenvalues obtained with the method of Saraceno and Vergini~\cite{Vergini1995} the eigenvalues of the billiards.}
        \label{Nfluc}
        \end{figure}

To better understand the structure of the spectra, we computed for each experimentally determined eigenvalue the associated wave function and Husimi function numerically using the method of Vergini and Saraceno~\cite{Vergini1995}. This method yields all eigenvalues and eigenfunctions in a narrow energy range by solving a generalized eigenvalue problem in terms of integrals over the boundary. In order to avoid the missing of eigenvalues we computed them in a sliding energy intervall with width of approximately 1/100 of the mean level spacing. Accordingly the resulting list contains replications of each eigenvalue. To select the eigenvalues of the billiards, we compared the fluctuating parts $N_{\rm fluc }(k)$ of the experimental (full black line) and the numerical (dashed red line) integrated resonance densities shown in Fig.~\ref{Nfluc} for B1. Note the slight shift between the curves. It is attributed to the shrinking of the microwave billiard when cooling it down to 4.2 Kelvin. To test whether pairs of eigenvalues are indeed doublet partners, i.e., that they do not differ due to the limited numerical accuracy, we computed the overlap of their wave functions. This provides a very efficient method for the extraction of the eigenvalues, because the overlap must vanish for genuine doublet states.

A Husimi function on the PSOS is defined as the projection of the normal derivative of the associated wave function at the boundary onto a coherent state~\cite{Backer2004a}. We use a periodization of a one-dimensional coherent state on the boundary of the billiard,
\begin{eqnarray}
{\rm H}_n(p,q)&=&\frac{1}{2\pi k_n}\frac{1}{\int_0^L{\rm d}q^\prime\left\vert\left\langle\hat n(q^\prime),\vec\nabla\Psi_n(q^\prime)\right\rangle\right\vert^2}\\
&\times&\left\vert\int_0^L{\rm d}q^\prime\left\langle\hat n(q^\prime),\vec\nabla\Psi_n(q^\prime)\right\rangle C^\rho_{(p,q)}(q^\prime;k_n)\right\vert^2\, ,
\nonumber
\end{eqnarray}
with
\begin{eqnarray}
&&C^\rho_{(p,q)}(q^\prime;k_n)=\left(\frac{k_n}{\pi\rho^2}\right)^{1/4}\\
&\times&\sum_{m=-\infty}^\infty\exp\left(ipk_n\left(q^\prime-q+mL\right)-\frac{k_n}{2\rho^2}\left(q^\prime-q+mL\right)^2\right)\, .\nonumber
\end{eqnarray}
Here, the quantity $\rho$ sets the resolution of the Husimi plots.
We also looked at the projection of the Husimi function onto the $p$ axis, i.e., its integral over the $q$ coordinate,
\begin{equation}
{\rm H}_n(p)=\int_0^L{\rm d}q{\rm H}_n(p,q).
\end{equation}

\begin{figure}[h!]
\centering
\begin{center}
\includegraphics[width=6cm]{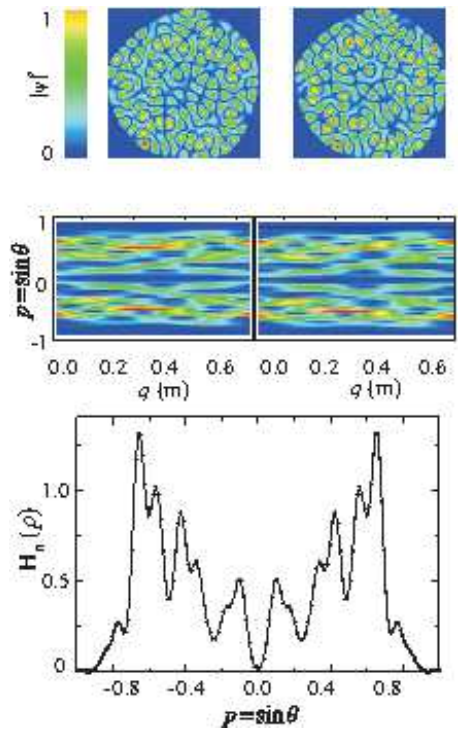}
\end{center}
\caption{(Color online) Squared modulus of the computed wave functions in the billiard plane (upper panels), the associated Husimi functions (middle panels) and the projection of the Husimi function of the left doublet partner onto the $q$ axis (lower panels) for B1, where the color scale is given to the left of the upper panels. Shown are the results for the chaotic modes with numbers 153 (left) and 154 (right). The wave functions are spread over the whole billiard area and the Husimi functions extend over the whole chaotic part of the PSOS. Likewise, the projection ${\rm H}_n(p)$ is nonvanishing for most values of $p$.}
\label{fig:chaos}
\end{figure}

\begin{figure*}[ht]
\begin{center}
\includegraphics[width=17 cm]{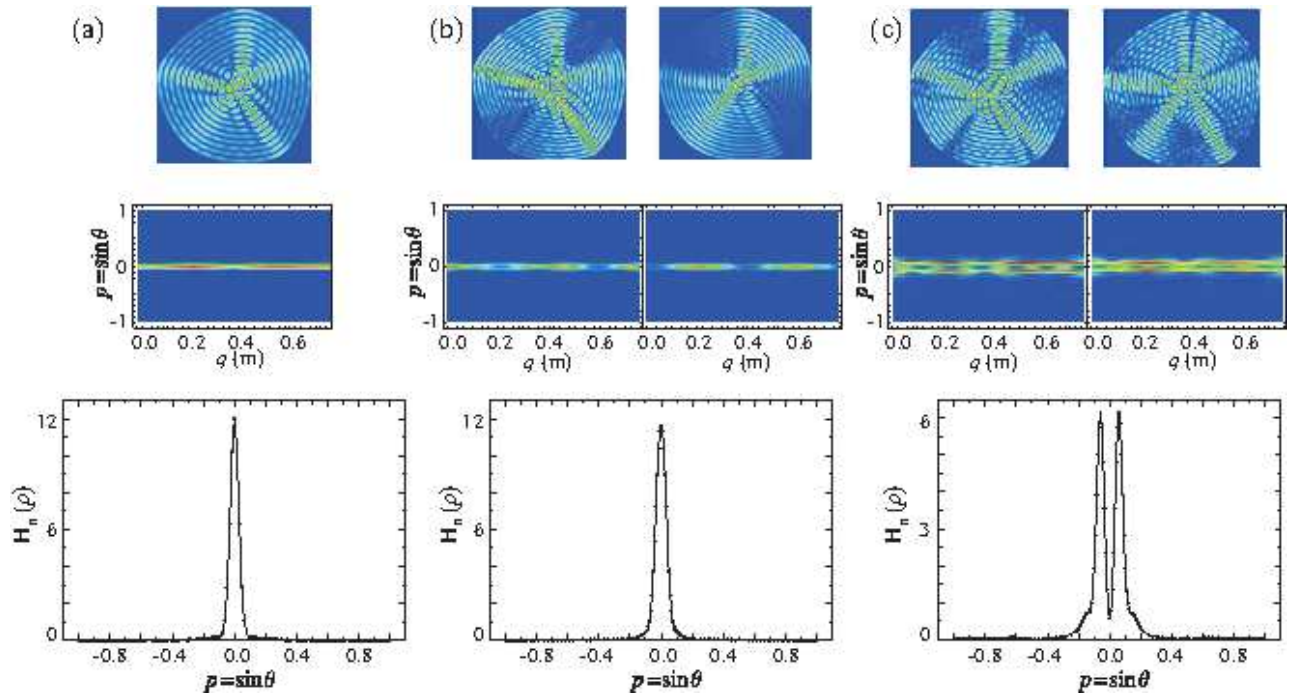}
\end{center}
\caption{(Color online) Same as Fig.~\ref{fig:chaos}. Panels (a) show the wave functions, Husimi functions and the projection of the Husimi function of the left doublet partner onto the $q$ axis for a singlet state of B1. Panels (b) present these functions for the pair of barrier KAM modes with numbers 575 and 576, panels (c) those for that with numbers 579 and 580. The wave functions are localized along the BBOs, the Husimi functions around the barrier of the KAM tori. Their projections exhibit a maximum along the $p=0$ line for the former pair, a mimimum that is bordered by two sharp maxima for the latter one.}
\label{fig:singlet_B1}
\end{figure*}

\begin{figure*}[ht]
\centering
\begin{center}
\includegraphics[width=17 cm]{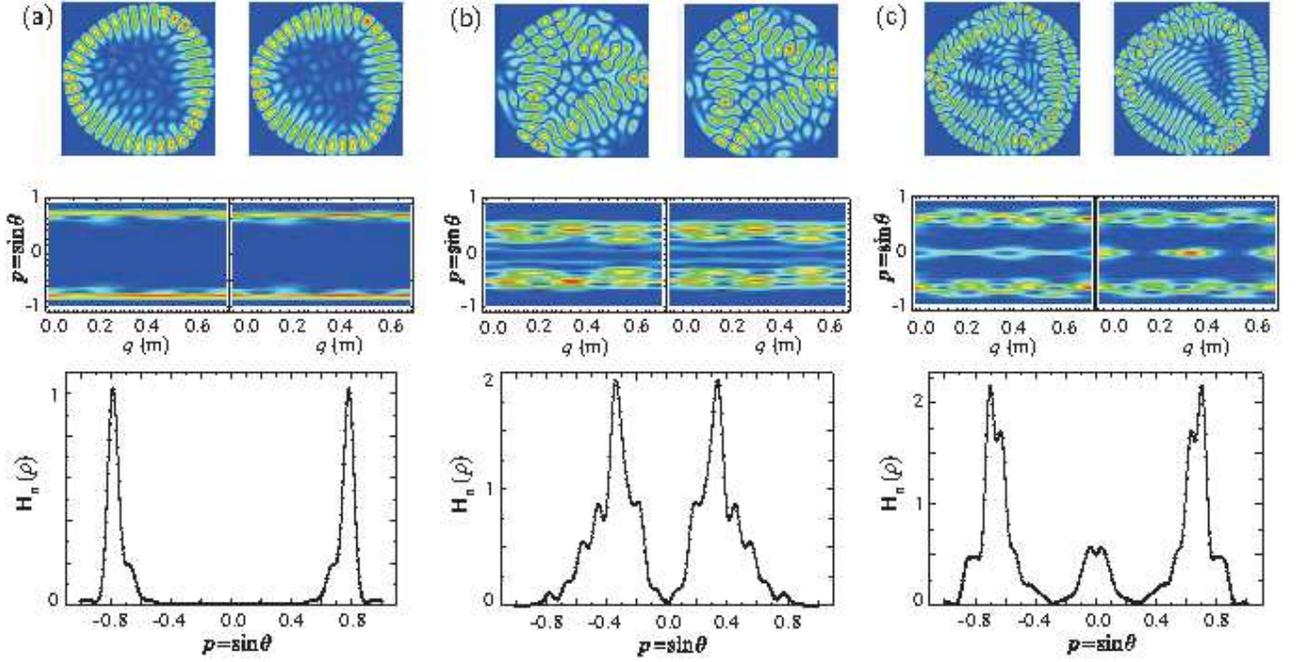}
\end{center}
\caption{(Color online) Same as Fig.~\ref{fig:chaos}. Panels (a) show the wave functions, Husimi functions and the projection of the Husimi function of the left doublet partner onto the $q$ axis for the doublet with mode numbers 113 (left) and 114 (right) of B1. The wave functions are localized close to the boundary and the Husimi functions near $p=\pm1$. Likewise, the projection ${\rm H}_n(p)$ is nonvanishing only there. Hence, these modes correspond to whispering gallery modes. In panels (b) these functions are depicted for the island modes with numbers 123 and 124. The wave functions are localized around the regular orbit shown in Fig.~\ref{fig:insel} and the Husimi functions around the islands in the chaotic sea also shown there. Panels (c) present those for the hybrid modes with numbers 259 and 260. The Husimi functions exhibit a high intensity at $p=0$ and at some value $\vert p\vert\lesssim 1$.}
\label{fig:whisp_B1}
\end{figure*}

For B1 we determined the wave functions and Husimi functions of 15 singlets and 394 doublet pairs. 
Most of the doublets are localized in the chaotic part of the PSOS. In Fig.~\ref{fig:chaos} we show for one example the wave functions (upper panels), the Husimi functions (middle panels) and the projection of the left Husimi function onto the $q$ axis (lower panel).
The associated Husimi functions extend over the whole chaotic part of the PSOS, thus they correspond to ``chaotic'' modes~\cite{Backer2004a}. In Fig.~\ref{fig:singlet_B1} (a) we show the functions for one of the singlet states. 
The wave function resembles a $J_0$-Bessel function which is deformed at the corners of the billiard. The associated Husimi function is shown in the middle panel of Fig.~\ref{fig:singlet_B1} (a). It is strongly localized around $p=0$ and its projection onto the $q$ axis shown in the lower panel of Fig.~\ref{fig:singlet_B1} (a) exhibits there a sharp peak. This demonstrates that the singlets indeed are associated with the BBOs in the classical billiard. Furthermore, there are modes that are essentially localized in the integrable part of the PSOS. These ``regular'' modes can be assigned to four different families. {\bf\emph{Barrier KAM modes}}: In Figs.~\ref{fig:singlet_B1} (b) and (c) two examples are presented. The wave functions are localized along the BBOs and likewise the Husimi functions in the region of the KAM tori barrier. The projection of the Husimi function shown in Fig.~\ref{fig:singlet_B1} (b) exhibits a sharp maximum at $p=0$, the one shown in Fig.~\ref{fig:singlet_B1} (c) has there a dip which is bordered by two sharp maxima. {\bf\emph{Whispering gallery modes}}: An example for such a doublet is shown in Fig.~\ref{fig:whisp_B1} (a). 
The wave functions are localized close to the billiard boundary, the Husimi functions close to $p=\pm 1$ and their projections onto the $q$ axis exhibit there sharp peaks. {\bf\emph{Island modes}}: In Fig.~\ref{fig:whisp_B1} (b) the pairs of wave functions and Husimi functions are depicted for one example. They are localized around the regular orbit associated with the regular islands in the chaotic sea presented in Fig.~\ref{fig:insel}. {\bf\emph{Hybrid modes}}: One example is shown in Fig.~\ref{fig:whisp_B1} (c). The associated Husimi functions are mostly localized in the region outside the KAM tori but also have a nonvanishing component at $p=0$.  

Similarly, we determined for B2 the wave functions and Husimi functions of $15$ singlets and $390$ doublet pairs. For B1 (B2) we found altogether $41$ ($13$) pairs of wave functions that correspond to whispering gallery modes. Amongst the KAM modes the Husimi functions of $21$ ($20$) pairs exhibit a maximum at $p=0$, while those of $24$ ($15$) pairs have a minimum there. Furthermore, we found $12$ ($31$) hybrid modes. Thus we found for B2 less ``regular'' modes than for B1. This is consistent with the fact that the regular regions in the PSOS of B2 are narrower than those for B1 (see Fig.~\ref{fig:whisp}). 

\section{\label{Fluct}Fluctuation properties in the spectrum of one doublet partner}
For the study of the spectral properties we separated the eigenvalue spectra obtained from the measured resonance spectra into $3$ parts, one part containing the smaller resonance frequencies $f^l_n=ck^l_n/(2\pi)$ of the doublets, i.e., that of the left resonances of the doublets in the transmission spectra (see Fig.~\ref{fig:spectra}), a second one comprising the larger 'right' ones denoted by $f^r_n=ck^r_n/(2\pi)$, and a third one containing the singlets. That this is possible is a clear indication that the correlations between the doublet partners and that of the doublets with the singlets are much weaker than those amongst the left or the right doublet partners. We indeed could simply disregard the singlets and consider the spectral properties of the eigenvalue sequences $\{k^l_n\}$ and $\{k^r_n\}$ separately. Since they are essentially identical on the scale of the mean eigenvalue spacing, we restrict in the following to those of $\{k^r_n\}$.  

\begin{figure}[h!]
        \centering
        \includegraphics[width=\linewidth]{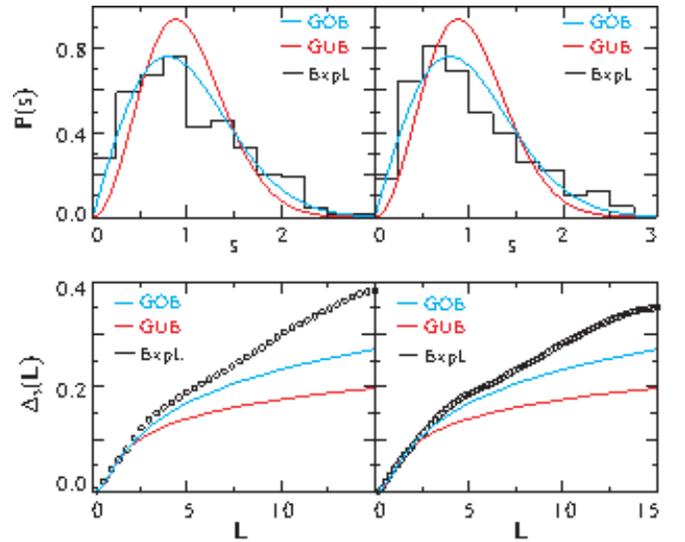}
        \caption{(Color online) Comparison of the spectral properties of billiards B1 and B2 (black) with those of random matrices from the GOE shown in blue (light gray) and the GUE depicted in red (dark gray). The upper left (right) panel shows the nearest-neighbor spacing distribution for billiard B1 (B2). Similarly, the lower panels show the $\Delta_3$ statistic.}
        \label{fig:vglgoegue}
\end{figure}

For the investigation of the spectral properties we first unfolded the spectra $\{k^l_n\}$ and $\{k^r_n\}$ with Weyl's formula~\cite{Weyl1912} to mean spacing unity, $\{k^l_n\}\to\{\tilde k^l_n\}$ and $\{k^r_n\}\to\{\tilde k^r_n\}$. Figure~\ref{fig:vglgoegue} shows in black the experimental results for the distribution ${\rm P}({\rm s})$ of the spacings ${\rm s}$ between adjacent eigenvalues and for the Dyson-Mehta statistic $\Delta_3({\rm L})$, which gives the least square deviation of the integrated resonance density of the unfolded eigenvalues from the straight line best fitting it in the interval of length ${\rm L}$~\cite{Mehta1990}. 
Due to the unidirectionality of the classical dynamics, i.e., the lack of ergodicity in the momentum part of the phase space~\cite{Gutkin2007}, the spectral statistics of the $\tilde k^r_n$ ($\tilde k^l_n$) with eigenstates localized in the chaotic part of the PSOS are predicted to coincide in the semiclassical limit with those of random matrices from the GUE~\cite{Mehta1990,Guhr1998}, even though the system is obviously time-reversal invariant~\cite{Leyvraz1996,Keating1997}. We, however, observe in Fig.~\ref{fig:vglgoegue} clear deviations of the experimental results from both the GUE curve shown in red (dark gray) and the GOE curve depicted in blue (light gray). 

Note that in each half of the PSOS the classical dynamics is not purely chaotic, but instead mixed regular / chaotic. Accordingly, we model the spectral properties using an ensemble of random matrices $H^{mixed}$, that are of the form~\cite{Lenz1990,Guhr1996}
\begin{equation}
H^{mixed}=\left({\begin{array}{cc}
                H\, &V_1\\
                V_1^\dagger\, &D_1
                \end{array}}
          \right)\, ,
\label{eq:mixedH}
\end{equation}
where $H$ and $D_1$ represent the chaotic and the regular part of the spectrum, respectively, and $V_1$ couples them. Here, $H$ is a GUE random matrix of the dimension $N_1\times N_1$ with variance $v_d=\sigma_H^2=\frac{1}{2N_1}$ for the diagonal elements and $v_n=\sigma_H^2/2$ for the off-diagonal ones. For this choice of the variances the eigenvalues take values from the interval $[-1,1]$. Furthermore, $D_1$ is a diagonal matrix of the dimension $N_2\times N_2$. Its entries are uncorrelated random numbers drawn from a Poisson process with values from the same range as the eigenvalues of $H$~\cite{Guhr1990a}. We chose $N_1=300$ which approximately equals the number of chaotic states. They were identified as explicated in~\refsec{Num}. Similarly, we set $N_2=100$ which is the approximate number of regular modes. The $N_2\times N_1$ dimensional matrix $V_1$ describes the mixing of the chaotic and the regular states. It contains Gaussian distributed elements with the variance $\sigma_{v1}^2=\left(d\tau_1\right)^2$, that is, the mixing, or equivalently, the coupling parameter $\tau_1$ is measured in units of the mean level spacing, which can be approximated as $d\simeq \frac{\pi}{2N}\sqrt{\frac{2N_1}{N}}$ at the center of the spectrum. Here $N=N_1+N_2$ is the dimension of $H^{mixed}$. 

We computed the $\Delta_3$ statistic for the eigenvalues of ensembles of $100$ random matrices of the form Eq.~(\ref{eq:mixedH}) and determined that value of $\tau_1$ for which the mean square deviation between the numerical and the experimental result is smallest. For this, the eigenvalues $E_l,\, l=1,\cdots ,N$ of $H^{mixed}$ were unfolded to mean spacing unity on the basis of the semicircle law for the GUE which yields for the integrated resonance density of the eigenvalues of $N\times N$-dimensional random matrices from the GUE $N^{GUE}(e)=\frac{N}{\pi}e\sqrt{1-e^2}+\frac{N}{2}+\frac{N}{\pi}\arcsin{(e)}$. Since the mixing of the regular and the chaotic modes is weak, we used a fit function of that form as ansatz for the integrated resonance density of the eigenvalues of $H^{mixed}$, 
\begin{equation}
N(E_l)=p_1\cdot E_l\cdot\sqrt{1-E_l^2}+\frac{N}{2}+p_2\cdot\arcsin(p_3\cdot E_l). 
\label{eq:unfolding}
\end{equation}
Here, the coefficients $p_i,\, i=1,2,3$ are the parameters, that were determined from a fit of this ansatz to the integrated resonance density of $H^{mixed}$. We came to the result that this function describes the latter very well. The unfolded eigenvalues $\tilde E_l$ were obtained by replacing $E_l$ by $\tilde E_l=N(E_l)$. 

\begin{figure}[h!]
        \centering
        \includegraphics[width=\linewidth]{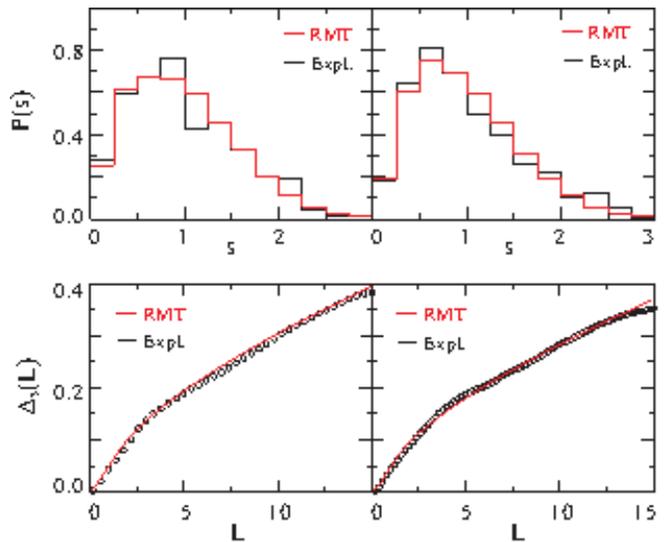}
        \caption{(Color online) Comparison of the spectral statistic of the measured eigenvalues shown in black with those of the RMT model Eq.~(\ref{eq:mixedH}) depicted in red (dark gray). The coupling parameter $\tau_1=0.135$ for B1 and $\tau_1=0.205$ for B2, respectively, was obtained from a fit of the $\Delta_3$ statistic deduced from the RMT model (\ref{eq:mixedH}) to the experimental result.}
        \label{fig:vglrmt}
         \end{figure}

The results for $\tau_1$ determined from the fit of the $\Delta_3$ statistic to the experimental ones are listed in the first column of~\reftab{table1}. 
\begin{table}[t!]
\caption{List of the values of the parameter $\tau_1$ which quantifies the mixing of the chaotic and the regular states in the RMT model Eq.~(\ref{eq:mixedH}). Its values were obtained from a fit of the $\Delta_3$ statistic deduced from the model to the experimental results. Here we chose the dimensions of $H$ and $D_1$ in Eq.~(\ref{eq:mixedH}) as $N_1=300,\, N_2=100$ and $N_1=2,\, N_2=1$, respectively.}
\label{table1}
\begin{tabular}{c|c|c}
\hline
\hline
   & $\tau_1$ & $\tau_1$\\
   & $N_1=300,\, N_2=100$ & $N_1=2,\, N_2=1$\\
\hline
B1 & 0.135 & 0.131 \\
B2 & 0.205 & 0.198 \\
\hline
\hline
\end{tabular}
\end{table}
Figure~\ref{fig:vglrmt} shows a comparison of the nearest-neighbor spacing distributions and the $\Delta_3$ statistics deduced from the experiments shown in black and the RMT model Eq.~(\ref{eq:mixedH}) depicted in red (dark gray), respectively. The agreement is very good. 

It is well known that the nearest-neighbor spacing distributions of the eigenvalues of random matrices from the Guassian ensembles are already well approximated by that of $2\times2$ matrices, the result being the renowned Wigner surmises. Accordingly, we set $N_1=2$ and $N_2=1$ in Eq.~(\ref{eq:mixedH}) and computed the nearest-neighbor spacing distributions for ensembles of $10000$ random matrices $H^{mixed}$. The values of $\tau_1$ determined from their fit to the expermental ones are listed in the second column of~\reftab{table1}. The corresponding distributions are barely distinguishable from those for $N_1=300,\, N_2=100$ depicted in red in Fig.~\ref{fig:vglrmt}, so we do not show them. We should note that it was demonstrated in Ref.~\cite{Baecker2011} that the nearest-neighbor spacing distribution of a system with a mixed phase space exhibits a fractional power law. This was attributed to varying couplings, i.e., dynamical tunneling rates between the different states in a regular region (see colored regions in Fig.~\ref{fig:psos}) and the chaotic sea. To resolve them we would need to measure resonance spectra up to much higher frequencies which would imply a loss of the analogy between the microwave billiard and the corresponding quantum billiard. The very good agreement between the experimental results for the spectral properties and those deduced from the RMT model Eq.~(\ref{eq:mixedH}) demonstrate that just one coupling parameter $\tau_1$ is sufficient to attain a good RMT description of the former. In the following three sections we investigate the splittings of the pairs of resonance frequencies, i.e., of the wave numbers $\delta_n=k^r_n-k^l_n$ of the doublet partners.  

\section{\label{Stat}Splitting-weighted density}
\begin{figure}[h!]
        \centering
        \includegraphics[width=\linewidth]{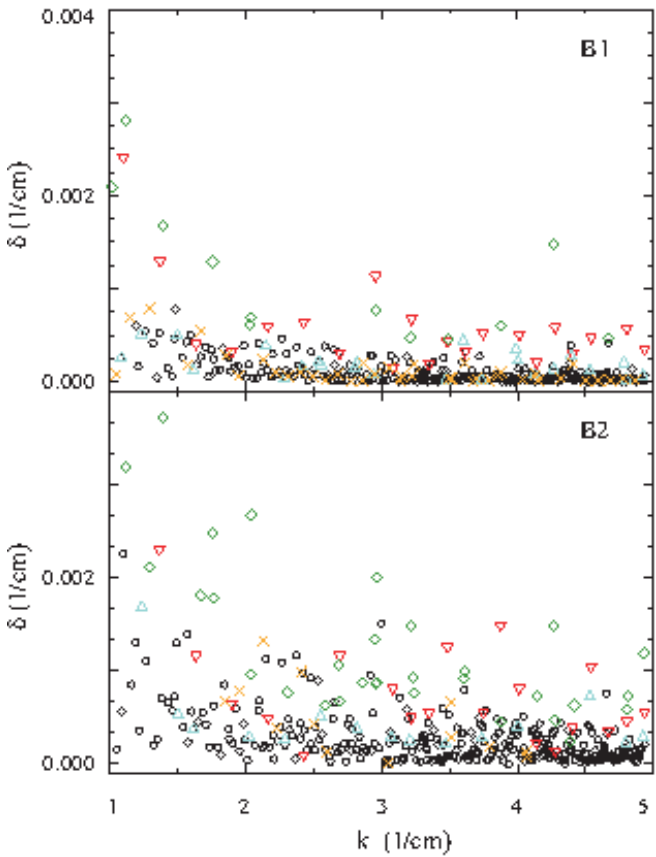}
        \caption{(Color online) Experimentally determined splittings, that is, differences $\delta=k^r_n-k^l_n$ for B1 (upper panel) and B2 (lower panel) versus $k=k^r_n$. Here, $k^r_n$ and $k^l_n$ denote the wave numbers of the right and the left doublet partners, respectively. The black circles mark the chaotic modes, the orange crosses whispering gallery modes (see panels (a) in Fig.~\ref{fig:whisp_B1}). Furthermore, the red downward and the  blue upward triangles mark the splittings of the KAM modes with a maximum and a minimum of the Husimi functions at $p=0$, respectively (see panels (b) and (c) in Fig.~\ref{fig:singlet_B1}). The green diamonds mark hybrid modes (see panels (c) of Fig.~\ref{fig:whisp_B1}).}
\label{fig:splitting_whisp_diag_WD}
\end{figure}

In Fig.~\ref{fig:splitting_whisp_diag_WD} the splittings $\delta_n=k^r_n-k^l_n$ of the doublet partners associated with chaotic states are marked by black circles. Those corresponding to whispering gallery modes (see Fig.~\ref{fig:whisp_B1} (a)), marked by orange crosses in Fig.~\ref{fig:splitting_whisp_diag_WD} have comparatively small splittings, i.e., tunneling rates. 
The eigenvalues of the doublet partners corresponding to KAM modes associated with Husimi functions that exhibit a maximum (minimum) at $p=0$ (see panels (b) and (c) in Fig.~\ref{fig:singlet_B1})) are marked by red downward (blue upward) triangles. They show large splittings. The remaining large splittings, marked as green diamonds in Fig.~\ref{fig:splitting_whisp_diag_WD}, are basically related to hybrid modes (see Fig.~\ref{fig:whisp_B1} (c)). 

In conclusion, the Husimi functions of the doublet partners corresponding to the largest splittings are \emph{all} nonvanishing around $p=0$, i.e., localized in the vicinity of the BBOs, while the other ones vanish there. This corroborates our assumption that the splittings are due to tunneling from the $p < 0$ part of the PSOS to the $p > 0$ one via the KAM tori region, and indicates that tunneling is enhanced due to scarring by the BBOs, as is confirmed at the end of this section. 

The splittings do not decrease exponentially with increasing wave number $k$, as might be expected in the semiclassical limit, but rather algebraically. 
Indeed, the average is well fitted by the function $\langle\delta\rangle\simeq n_1/k^{n_0}$, where the fit parameters equal
$n_0=1.5603,\, n_1=0.0007$ for B1 and $n_0=1.2907,\, n_1=0.0015$ for B2.
This decay behavior can be attributed to the fact that the PSOS consists of regular and chaotic parts. The associated splittings decay exponentially with different rates. We consider all these contributions and their sum yields the algebraic decay deduced from the data in Fig.~\ref{fig:splitting_whisp_diag_WD}~\cite{Dittes2000}. 

\begin{figure}[tb]
        \centering
        \includegraphics[width=\linewidth]{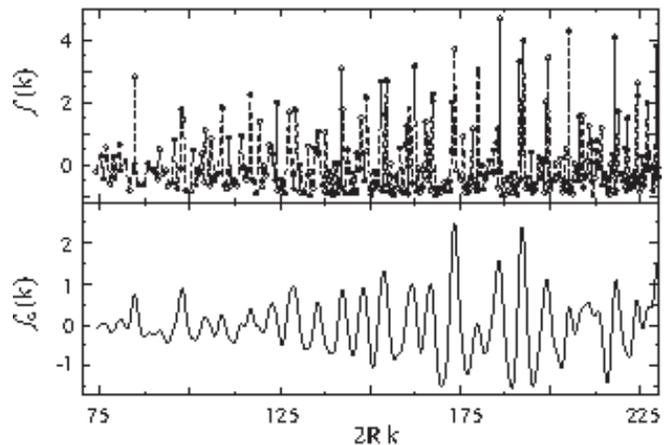}
        \caption{Experimentally determined splitting-weighted density of states $f(k)$ (upper panel) (see Eq.~(\ref{eq:2}))
        and the result for its convolution $f_G(k)$ with a Gaussian of width $\sigma_G$ (see Eq.~(\ref{eq:3})) with 
        $\sigma_G=1.0$ (lower panel) for B1. Large splittings seem to occur periodically.}
        \label{fig:splittingwheighted1}
\end{figure}

\begin{figure}[tb]
        \centering
        \includegraphics[width=\linewidth]{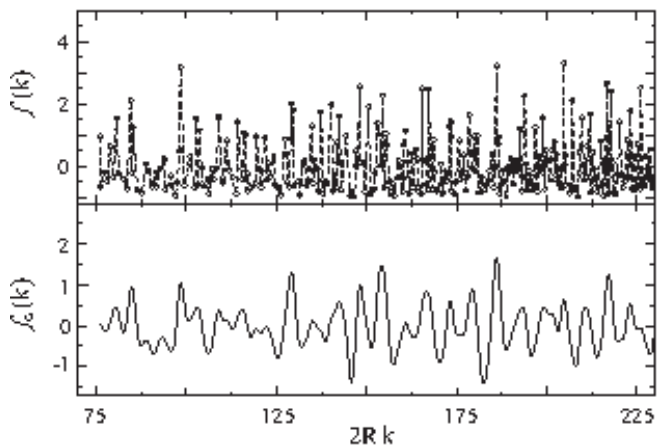}
        \caption{Same as Fig.~\ref{fig:splittingwheighted1} but for B2.}
        \label{fig:splittingwheighted2}
\end{figure}

\begin{figure}[tb]
        \centering
        \includegraphics[width=\linewidth]{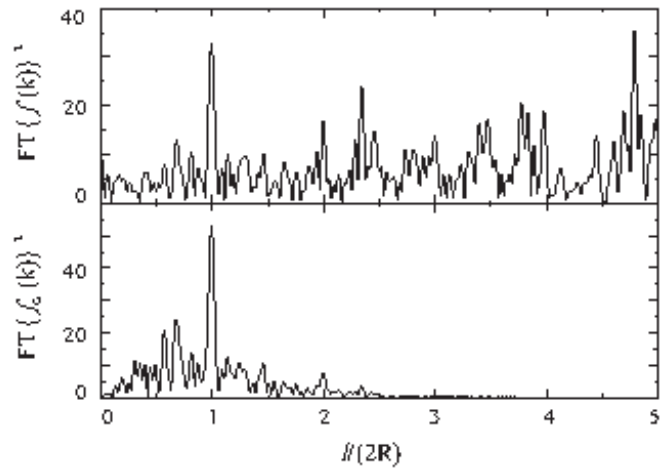}
        \caption{Comparison of the Fourier transforms of the splitting-weighted density of states $f(k)$
        (upper panel) with that of $f_G(k)$ for 
        $\sigma_G=1.0$ (lower panel) for B1 where $f(k)$ and $f_G(k)$ are shown in Fig.~\ref{fig:splittingwheighted1}. Both exhibit a dominant peak at the length of the diameter.}
        \label{fig:fftsplittingwheighted1}
\end{figure}

\begin{figure}[tb]
        \centering
        \includegraphics[width=\linewidth]{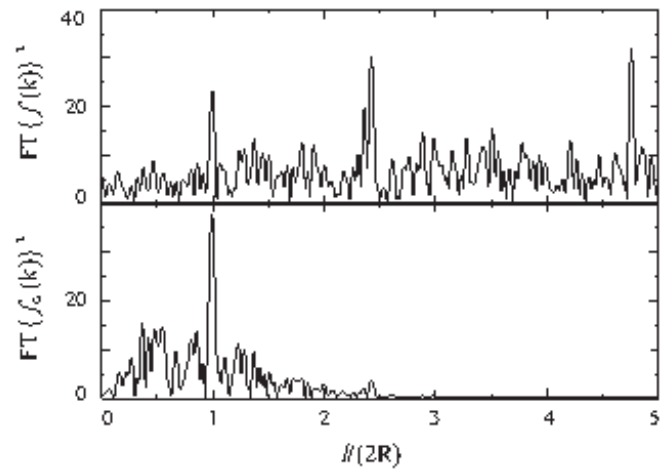}
        \caption{Same as Fig.~\ref{fig:fftsplittingwheighted1} but for B2.}
        \label{fig:fftsplittingwheighted2}
\end{figure}

Following Ref.~\cite{Creagh1996} we evaluated the splitting-weighted density of states, which is defined as
\begin{equation}
f(k)=\sum_n\left(\frac{\delta_n}{\langle\delta\rangle}-1\right)\delta (k-k_n).
\label{eq:2}
\end{equation}
The results are shown for B1 and B2 in the upper panels of Figs.~\ref{fig:splittingwheighted1} and~\ref{fig:splittingwheighted2}, respectively.
%
%
In distinction to the data shown in Fig.~\ref{fig:splitting_whisp_diag_WD} here the $k$-dependent average was subtracted from the splittings and the result was divided by it, thus yielding the fluctuations of the splittings in units of their average value. For a better understanding of the oscillatory structure we computed the smoothed function $f(k)$, which is obtained from the convolution of the splitting-weighted density of states with a Gaussian of width $\sigma_G$, 
\begin{equation}
f_G(k)=\frac{1}{\sqrt{2\pi}\sigma_G}\int_0^\infty{\rm d}Kf(K)e^{-\frac{(k-K)^2}{2\sigma_G^2}}.
\label{eq:3}
\end{equation}
A periodic recurrence of the largest splittings is revealed for both billiards (see lower panels of Figs.~\ref{fig:splittingwheighted1} and~\ref{fig:splittingwheighted2}). To determine the periods we computed the Fourier transforms of $f(k)$ and $f_G(k)$ that are shown in Figs.~\ref{fig:fftsplittingwheighted1} and~\ref{fig:fftsplittingwheighted2}. Especially the Fourier transform of the smoothed functions reveal that the dominant period for the recurrence of the largest splittings equals the length of the diameter. Actually, the high peaks occurring in the Fourier transform of $f(k)$ at $l/(2R)\simeq 2.3$ and $4.8$ shown in the upper panels of Figs.~\ref{fig:fftsplittingwheighted1} and~\ref{fig:fftsplittingwheighted2} are remnants from regular periodic orbits, that show up at these lengths in the corresponding length spectra, that is, the Fourier transform of the fluctuating part of the resonance density itself, as exceptionally high peaks. They fade away rapidly with increasing width $\sigma_G$ when computing the convolution Eq.~(\ref{eq:3}) and completely disappeared in the Fourier transform shown in the lower panels of Figs.~\ref{fig:fftsplittingwheighted1} and~\ref{fig:fftsplittingwheighted2}. These observations corroborate our above assumption, deduced from the observation that the splittings are large for modes with nonvanishing Husimi functions in the KAM tori region around the BBOs, that tunneling is enhanced through a scarring by the BBOs~\cite{Wilkinson1996}. Actually, in Ref.~\cite{Creagh1996} the splitting-weighted density of states Eq.~(\ref{eq:2}) could be expressed in terms of a trace formula for complex orbits which tunnel through a potential barrier. It is still a challenging problem to determine such complex orbits for dynamical tunneling but beyond the scope of the present paper as it would require independent studies with the focus on the semiclassical rather than the RMT approach. 
%
\section{\label{RMT}Random matrix model for the splitting distribution}
We demonstrated in \refsec{Num} that the spectral properties of B1 and B2 are universal and well described by RMT. Accordingly we developed a RMT model for the statistical properties of the splittings. The results of the previous sections suggest that the tunneling between the states corresponding to clockwise and anticlockwise motion, respectively, is due to their overlap with a singlet state.  Therefore, we  used a RMT model of the form~\cite{Bohigas1993,Leyvraz1996a,Schlagheck2006} 
\begin{equation}
H^{splitting}=\left({\begin{array}{ccc}
        H^{mixed}\, &V\, &0\\
        V^\dagger\, &D_0\, &V^\dagger\\
        0\, &V\, & {H^*}^{mixed}\\
        \end{array}}
  \right)\, .
                   \label{eq:1}
\end{equation}
\noindent
Here, the $N\times N$ matrices $H^{mixed}$ and ${H^*}^{mixed}$ are associated with the parts of the Hilbert space corresponding to clockwise and anticlockwise motion, respectively. These matrices were introduced in Eq.~(\ref{eq:mixedH}) to account for the mixed regular / chaotic doublet states. The matrix $D_0$ is diagonal with uncorrelated random numbers drawn from a Poisson process as entries and represents the singlet states. Its dimension is set to unity and its entry $E_R$ takes a value from the interval $[-1,1]$ so the spectra of $H^{mixed}$ and $D_0$ have the same range of values. The tunneling from the upper to the lower part of the PSOS, or equivalently, the splittings into doublets of nearly degenerate eigenvalues is induced by coupling $H^{mixed}$ and ${H^*}^{mixed}$ via the singlet state. Due to time-reversal invariance the coupling to the latter is the same for $H^{mixed}$ and ${H^*}^{mixed}$ and thus caused by the same $N$-dimensional vector $V$. Its entries are Gaussian distributed random numbers with variance $\sigma_{v2}^2=\left(d\tau_2\right)^2\sigma_H^2$. Here, $\sigma_H^2$ equals the variance $v_n$ of the off-diagonal elements of $H$ in Eq.(\ref{eq:mixedH}). 

The splitting, or tunneling parameter $\tau_2$ measures the strength of the coupling in units of the mean spacing $d$ of the eigenvalues of $H^{mixed}$. It was determined from a fit of the splitting distribution deduced from the RMT model Eq.~(\ref{eq:1}) to the experimental one.
For this we first identified the eigenvalue $E_R^s$ corresponding to the singlet and those of the doublets. Here we used the fact that the sizes of the splittings of the experimental doublets are much smaller than the mean spacing between adjacent left and right doublet partners, respectively (see Fig.~\ref{fig:spectra}), that is, the coupling between $H^{mixed}$ and ${H^\star}^{mixed}$ was expected to be very weak. Furthermore, the eigenvalue $E_R^s$ in general coincided with that eigenvalue of $H^{splitting}$ which was closest to $E_R$. In order to obtain $\tau_2$ in units of the mean level spacing $d$ we computed the splittings using the unfolded eigenvalues of $H^{splitting}$. As the latter essentially coincide with the eigenvalues of $H^{mixed}$, viz., ${H^*}^{mixed}$ (see below) we could use Eq.~(\ref{eq:unfolding}) for their unfolding. Similarly, the experimental splitting distributions were obtained from the splittings of the unfolded left and right doublet partners, $\tilde\delta =\tilde k^r_n-\tilde k^l_n$, respectively. 

\begin{figure}[tb]
        \centering
        \includegraphics[width=\linewidth]{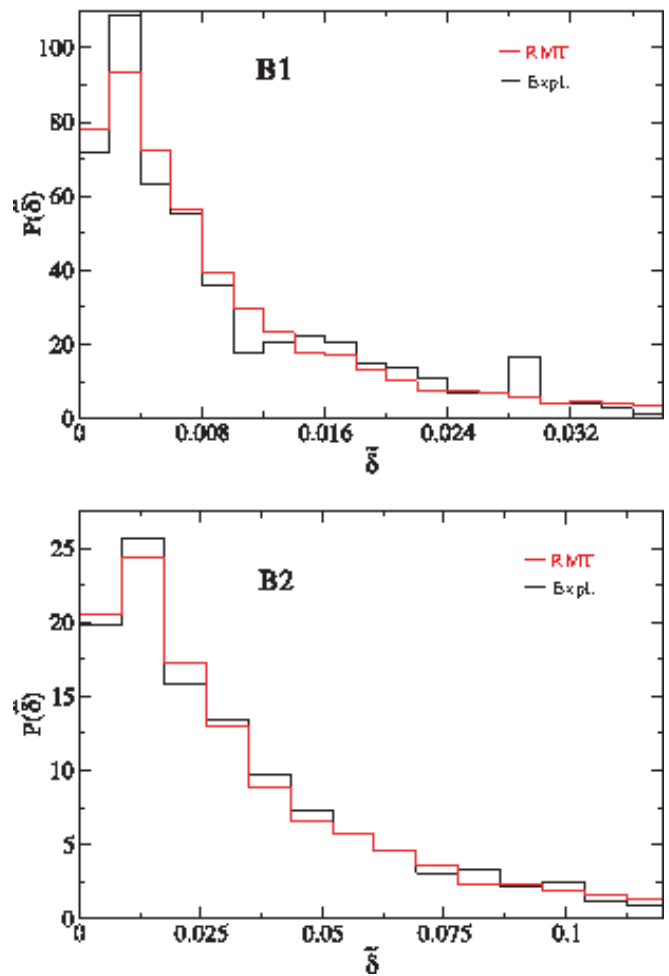}
        \caption{(Color online) Comparison of the experimental splitting distributions shown in black with those obtained from the RMT model Eq.~(\ref{eq:1}) depicted in red (dark gray), with $H^{mixed}$ replaced by a $N\times N$-dimensional random matrix from the GUE with $N=300$. That for B1 is shown in the upper panel, that for B2 in the lower one. The splittings $\tilde\delta$ were obtained by first unfolding the sequences of left and right doublet partners, respectively, to mean spacing one.}
\label{fig:splitting_distr_exp_rmt}
\end{figure}

We computed the splitting distributions of ensembles of $100$ random matrices $H^{splitting}$ and determined $\tau_2$ from a fit to the experimental one. In Fig.~\ref{fig:splitting_distr_exp_rmt} we compare the experimental splitting distribution shown in black to that obtained from the RMT model depicted in red (dark gray). 
The resulting values of $\tau_2$ are listed for B1 and B2 in the first column of~\reftab{table2}. 
\begin{table}[h!]
\caption{List of the values of the tunneling parameter $\tau_2$ accounting for the splitting of the eigenvalues of the RMT model Eq.~(\ref{eq:1}) into nearly degenerate doublet states. Its values were obtained from a fit of the splitting distribution deduced from the model to the experimental results. Here we replaced $H^{mixed}$ in Eq.~(\ref{eq:1}) by a $N\times N$-dimensional random matrix from the GUE with $N=300$ and $N=2$, respectively. Also listed are the results for the fit of the truncated Cauchy distribution to the experimental splitting distributions.}
\label{table2}
\begin{tabular}{c|c|c|c}
\hline
\hline
   & $\tau_2$ & $\tau_2$ & $\tau_2$\\
   & $N_1=300$ & $N_1=2$ & Cauchy\\
\hline
B1 & 0.093 & 0.094 &0.089 \\
B2 & 0.183 & 0.185 &0.182\\
\hline
\hline
\end{tabular}
\end{table}
As expected $\tau_2$ is larger for B2 than for B1, since the regular region around the BBOs (see Fig.~\ref{fig:whisp}) is narrower for B2. 

Recall that the specific form~(\ref{eq:mixedH}) of $H^{mixed}$ is justified by the fact that the upper and the lower parts of the PSOS exhibit a mixed regular / chaotic dynamics. However, since the chaotic component dominates, we may expect that the regular part $D_1$  of $H^{mixed}$ plays no essential role for the splitting distribution. To check this assumption we evaluated numerically  the splitting distribution for the RMT model~(\ref{eq:1}) with  $H^{mixed}$  drawn from two different ensembles. In the first case we chose $H^{mixed}$ as in Eq.~(\ref{eq:mixedH}) and fixed $\tau_1$ as listed in~\reftab{table1} while in the second one $H^{mixed}$ was drawn from the GUE, i.e., we set $N_2=0$. We arrived at the conclusion that the shape of the splitting distribution and also the values of $\tau_2$ determined from its fit to the experimental ones was the same for both ensembles. Likewise, the experimental splitting distributions barely change when we consider only doublets with the eigenvalues corresponding to chaotic states, i.e., when skipping those with a Husimi distribution localized close to $p=\pm 1$ or $p=0$ and around the small regular islands (see Figs.~\ref{fig:singlet_B1} (b), (c) and \ref{fig:whisp_B1} (a), (b)). We thus may conclude that the splitting distribution is solely determined by the parameter $\tau_2$ and does not depend on the classical dynamics in the upper and the lower half of the PSOS, respectively, as long as the chaotic component dominates. Actually, the red (dark gray) curves shown in Fig.~\ref{fig:splitting_distr_exp_rmt} were generated for $H^{mixed}$ drawn from the GUE, i.e., for $N_2=0$. We already mentioned above that we were not able to resolve tunneling rates corresponding to the individual states in the regular regions~\cite{Backer2010} within the experimentally accessible frequency range. The good agreement between the RMT model and the experimental results, however, corroborates the legitimacy of this ansatz. 

The considered RMT model, in fact, can be even further simplified because, similar to the spectral properties of the eigenvalues of $H^{mixed}$ the splitting distribution is well approximated by that of an ensemble of $2\times2$ random matrix from the GUE. Accordingly, we set $N=2$ in Eq.~(\ref{eq:1}) and again determined $\tau_2$ from a fit of the splitting distributions computed for ensembles of 10000 random matrices to the experimental ones. This yielded the values of the tunneling parameter $\tau_2$ listed in the second column of~\reftab{table2}. They are in good agreement with those obtained above for high-dimensional random matrices.
\begin{figure}[tb]
        \centering
       \includegraphics[width=\linewidth]{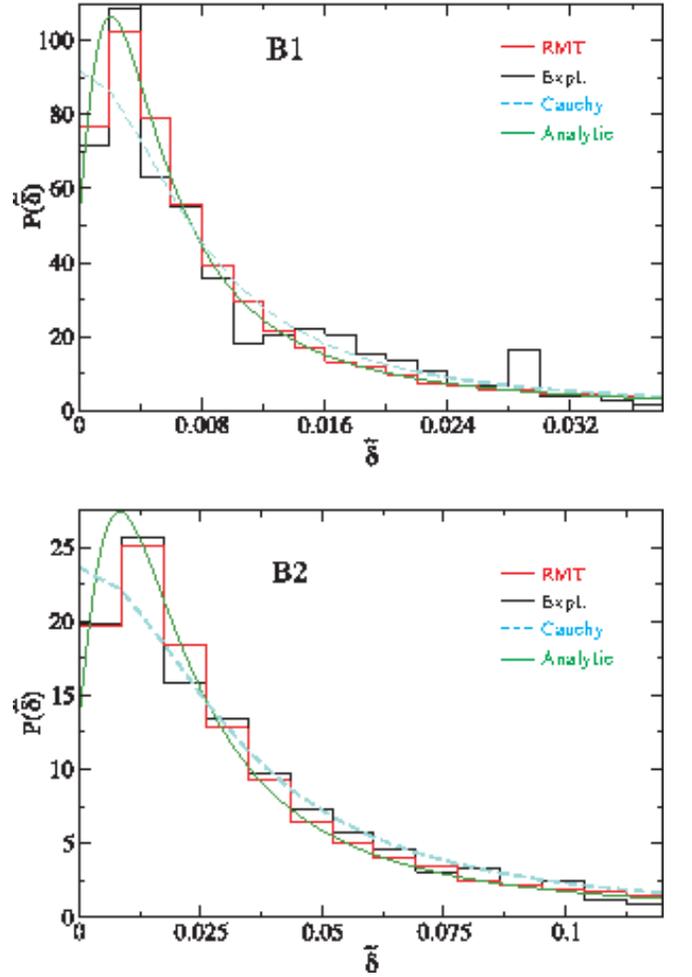}
        \caption{(Color online) Comparison of the experimentally determined splitting distributions shown as black histogram with that obtained from the RMT model Eq.~(\ref{eq:1}) depicted as red (dark gray) histogram, with $H^{mixed}$ replaced by a $2\times 2$ random matrix from the GUE. The green (gray) full curve shows the respective analytical result Eq.~(\ref{spl_distr}). The dashed (cyan) curve results from a fit of the truncated Cauchy distribution given in Eq.~(\ref{eq:cauchy}) to the experimental result. The experimental and the numerical splittings $\tilde\delta$ were obtained by first unfolding the sequences of left and right doublet partners, respectively, to mean spacing one.}
\label{figSPmodel}
\end{figure}
In Fig.~\ref{figSPmodel} we compare the resulting splitting distribution shown in red (dark gray) with the experimental one plotted in black. Again the agreement is very good.

\section{\label{Analytic}Analytical result for the splitting distribution}
In the following we derive an analytical expression for the splitting distribution based on the assumption that $H^{mixed}$ is a member of the GUE. The starting point is the equation for the eigenvalues of $H^{splitting}$. To obtain it we define the unitary matrix
\begin{equation}
\mathcal{U}=\left({\begin{array}{ccc}
        U\, &0\, &0\\
        0\, &\II\, &0\\
        0\, &0\, &U^\star\\
        \end{array}}
  \right)\, .
                   \label{eq:U}
\end{equation}
where $U$ is the matrix, that diagonalizes $H^{mixed}$,
\begin{equation}
H^{mixed}=U\hat EU^\dagger
\label{eq:UD}
\end{equation}
and $\hat E$ is a diagonal matrix that contains the eigenvalues $E_l,\, l=1,\cdots ,N$ of $H^{mixed}$, viz., ${H^\star}^{mixed}$. The unitary transformation $\tilde H^{splitting}=\mathcal{U}^\dagger H^{splitting}\mathcal{U}$ brings the matrix~(\ref{eq:1}) to the form
\begin{equation}
\tilde H^{splitting}
=\left({\begin{array}{ccc}
        \hat E\, &W\, &0\\
        W^\dagger\, &E_R\, &\tilde W^\dagger\\
        0\, &\tilde W\, & \hat E\\
        \end{array}}
  \right)\, ,
                   \label{eq:HD}
\end{equation}
where $W,\, \tilde W$ are $N$-dimensional vector defined as
\begin{equation}
W=U^\dagger V,\, \tilde W=U^T V\, ,
\end{equation}
and $E_R$ is the only matrix element of $D_0$. The eigenvalue equation for $\tilde H^{splitting}$,  
\begin{equation}
\det\left(\tilde H^{splitting}-\lambda\II\right)=0
\end{equation}
leads with $\mathcal{W}_i=\vert W_i\vert^2+\vert\tilde W_i\vert^2$ to the equation
\begin{equation}
(\lambda -E_R)=\sum_{i=1}^N\frac{\mathcal{W}_i}{\lambda -E_i}\, . 
\label{Eq:EV}
\end{equation}
This is a polynomial equation of order $N+1$ with the solutions yielding $N+1$ eigenvalues of $H^{splitting}$. The remaining $N$ eigenvalues coincide with those of $H^{mixed}$, viz., ${H^\star}^{mixed}$, i.e., with $E_l,\, l=1,\cdots ,N$. Thus the eigenvalue spectrum consists of $N$ doublets $(E_l,\, E_l^s=E_l+\Delta_l)$ and $1$ singlet $E_R^s=E_R+\Delta_R$. To obtain a conditional equation for the shift $\Delta_R$ we insert the ansatz $E_R^s=E_R+\Delta_R$ into Eq.~(\ref{Eq:EV}). This yields  
\begin{equation}
\Delta_R=\sum_{i=1}^N\frac{\mathcal{W}_i}{(E_R-E_i)+\Delta_R}\, . 
\label{Eq:DR0}
\end{equation}
We observed in our numerical simulations, that generally, the differences $E_R-E_l$ between the singlet eigenvalue $E_R$ and the eigenvalues of $H^{mixed}$ are large in comparison to $\Delta_R$ so
\begin{equation}
\Delta_R\simeq\sum_{i=1}^N\frac{\mathcal{W}_i}{(E_R-E_i)}\, . 
\label{Eq:DR}
\end{equation}
Note, however, that this must not always be the case, because $E_R$ and the eigenvalues $E_l,\, l=1,\cdots ,N$ stem from independent spectra~\cite{Leyvraz1996a}. Actually, Eq.~(\ref{Eq:DR}) coincides with Eq.~(3.2) of Ref.~\cite{Leyvraz1996a} derived for the splitting of a regular state caused by chaos-assisted tunneling. 

To account for the dynamical tunneling between two chaotic regions via a regular one we focus on the splittings $\Delta_l$. Inserting the ansatz $E_l^s=E_l+\Delta_l$ into Eq.~(\ref{Eq:EV}) yields for the splitting of the $l$-th doublet 
\begin{equation}
E_l+\Delta_l-E_R=\sum_{{i=1}\atop{i\ne l}}^N\frac{\mathcal{W}_i}{(E_l-E_i)+\Delta_l}+\frac{\mathcal{W}_l}{\Delta_l}\, . 
\label{Eq:DL}
\end{equation}
Without loss of generality we may set $E_R$ equal to zero, or equivalently, shift the origin of the eigenvalues $E_l\to E_l-E_R$ and thus obtain an equation for the splittings which only depends on the eigenvalues of the Hamiltonian of the uncoupled system and on the coupling matrix elements $\mathcal{W}_l$, i.e., on the variance of the tunneling matrix, 
\begin{equation}
E_l+\Delta_l=\sum_{{i=1}\atop{i\ne l}}^N\frac{\mathcal{W}_i}{(E_l-E_i)+\Delta_l}+\frac{\mathcal{W}_l}{\Delta_l}\, .
\label{Eq:DLF}
\end{equation}
Based on this equation we derived an analytical expression for the splitting distribution. To further simplify this problem we exploited the result of the previous section, that the splitting distribution can be evaluated by using an ensemble of random $2\times 2$ matrices $H^{mixed}$ from the GUE. Accordingly, we set $N=2$ in Eq.~(\ref{Eq:DLF}). This yields, e.g., for $E_1$ 
\begin{equation}
E_1+\Delta_1=\frac{\mathcal{W}_2}{(E_1-E_2)+\Delta_1}+\frac{\mathcal{W}_1}{\Delta_1}\, .
\label{Eq:DLF2}
\end{equation}
Using the fact that the eigenvalues $E_l$ and also the spacings between them are generically much larger than the splittings,  $E_l\gg\Delta_l$ and $\vert E_{l+1}-E_l\vert\gg\Delta_l$, we obtain
\begin{equation}
\Delta_l\simeq\frac{\mathcal{W}_l}{\vert E_l\vert}\, ,\, l=1,2\, .
\label{DLF2A}
\end{equation} 
We checked numerically that this equation indeed provides a good approximation for Eq.~(\ref{Eq:DLF2}). The derivation of the analytical expression for the splitting distribution of the eigenvalues of $H^{splitting}$ defined in Eq.~(\ref{eq:1}), which is given by
\begin{equation}
{\rm P}(\Delta)=\frac{2}{\sqrt{\pi}}\frac{\tau_2^2}{(\Delta+\tau_2^2)^2}e^{-4\frac{\tau_2^4}{(\Delta+\tau_2^2)^2}}\left[1+8\frac{\tau_2^4}{(\Delta+\tau_2^2)^2}\right],
\label{spl_distr}
\end{equation}
is outlined in the appendix. We plotted in Fig.~\ref{figSPmodel} the curves resulting from this analytical expression (green full line) together with the experimental splitting distributions (black histogram) and the random-matrix simulations depicted as red (dark gray) histograms.

Interestingly nearly the same values are obtained for the tunneling parameter $\tau_2$ from a fit of the truncated Cauchy distribution~\cite{Leyvraz1996a,Schlagheck2006}
\begin{equation}
{\rm P}(\Delta)=\frac{2}{\pi}\frac{\tau_2^2}{\Delta^2+\tau_2^4}
\label{eq:cauchy}
\end{equation}
to the experimental results. The best fits are shown as dashed (cyan) curves in Fig.~\ref{figSPmodel} and the resulting values for $\tau_2$ are listed in the third column of~\reftab{table2}. Large deviations between both distributions are mainly observed around $\delta=0$. This similarity provides a link between the tunneling parameter $\tau_2$ and the tunneling matrix element investigated in the context of chaos-assisted tunneling in~\cite{Bohigas1993,Leyvraz1996a,Schlagheck2006,Mouchet2001,Mouchet2003} and expressed there entirely in terms of classical quantities. Note that, because of the complex structure of the PSOS of the constant-width billiards, this would be hard to achieve for their splitting distribution and tunneling rates.

\section{\label{Concl}Conclusions}
We investigated both experimentally and theoretically the spectral properties and the dynamical tunneling occurring in constant-width billiards with smooth boundaries. To this end we determined with unprecedented accuracy the resonance frequencies of two superconducting microwave billiards. A special dynamical feature of constant-width billiards is the unidirectionality, that is, the change of the direction of motion from clockwise (upper half of the PSOS) to anticlockwise (lower half of the PSOS) is not possible. In both parts of the PSOS the classical dynamics are predominantly chaotic. They are separated by a barrier of KAM tori. Although the transition through this barrier is forbidden in the classical system, in the corresponding quantum billiard such processes occur due to dynamical tunneling. They manifest themselves in a splitting of the vast majority of eigenstates into doublets of nearly-degenerate ones. In addition, there are a few singlet eigenstates localized on the BBOs that divide the PSOS into the parts of clockwise and anticlockwise motion. 

To study the spectral properties the doublets were decomposed into two sequences, one containing the smaller, the other one the larger doublet partners. We demonstrated, that the spectral properties are well described by those of random matrices of the form Eq.~(\ref{eq:mixedH}). These consisted of a GUE matrix modeling the chaotic part, which is coupled to a diagonal matrix containing uncorrelated random numbers and accounting for the regular part of the classical dynamics. Importantly the chaotic component is not modeled by GOE matrix, as might be  expected for systems with  time reversal invariance, due  to the unidirectional character of the billiard dynamics.

Furthermore, we investigated the effects of dynamical tunneling on the distribution of the splittings between the doublet partners. As opposed to the well studied problem of chaos-assisted tunneling, in the case of billiards of constant width the particle tunnels from one chaotic region of the PSOS to another one through a barrier of KAM tori. Using the experimentally determined eigenvalues of the constant-width billiards we computed the associated wave functions and Husimi functions. With this information at hand we could demonstrate that the splittings indeed are enhanced for states that are strongly localized  in the barrier region. We showed that the experimental splitting distributions are well described by the ensemble of random matrices~(\ref{eq:1}), where two complex conjugate GUE matrices are coupled via a single state. Based on this model we derived an analytical expression for the splitting distribution. We believe that this result is generally valid for the description of dynamical tunneling between two regions with chaotic dynamics with violated time-reversal invariance. Remarkably, the resulting formula for the splitting distribution depends only on one parameter that provides information on the strength of the tunneling. It would be of interest to know whether it can be expressed in terms of quantities characterizing the corresponding classical dynamics, e.g., by complex periodic trajectories~\cite{Creagh1996}. 

\begin{acknowledgments}
This work was supported by the DFG within the Collaborative Research Centers 634 and TR12 as well as within the research grant Gu 1208/1-1 (BG). We thank Birgit Quast who performed preliminary experiments with constant-widths billiards in the framework of her Diploma thesis. We are grateful in particular to S. L\"ock and R. Ketzmerick, and also to A. B\"acker, S. Creagh and S. Tomsovic for instructive and stimulating discussions. Moreover, BD acknowledges fruitful conversations with S. Fishman, F. Leyvraz and M. Horvat. 
\end{acknowledgments}
\appendix
\section{\label{Appendix}Derivation of the splitting distribution for $N=2$}
\begin{figure}[h!]
        \centering
        \includegraphics[width=\linewidth]{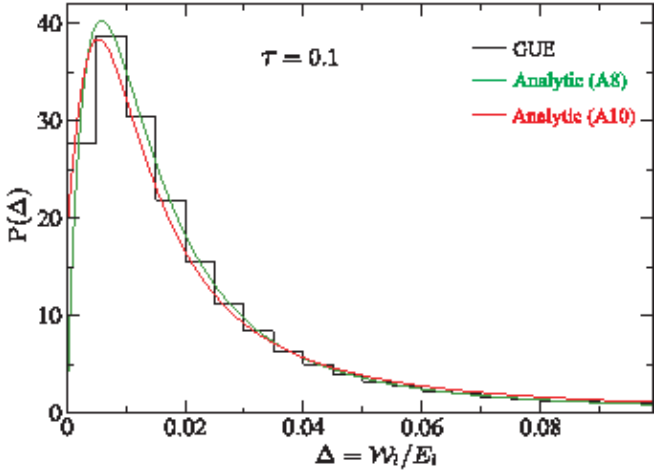}
        \caption{(Color online) Comparison of the distribution of the ratios $\mathcal{W}_l/E_l$ obtained for random matrices of the form Eq.~(\ref{eq:1}) with $H^{mixed}$ replaced by a random $2\times 2$ matrix from the GUE (black histogram) with the analytical results Eq.~(\ref{PDrsk}) depicted as green (gray) full line and Eq.~(\ref{PDappr}) shown as red (dark gray) histogram).}
\label{ev_verteilung}
\end{figure}

\begin{figure}[bt]
        \centering
        \includegraphics[width=\linewidth]{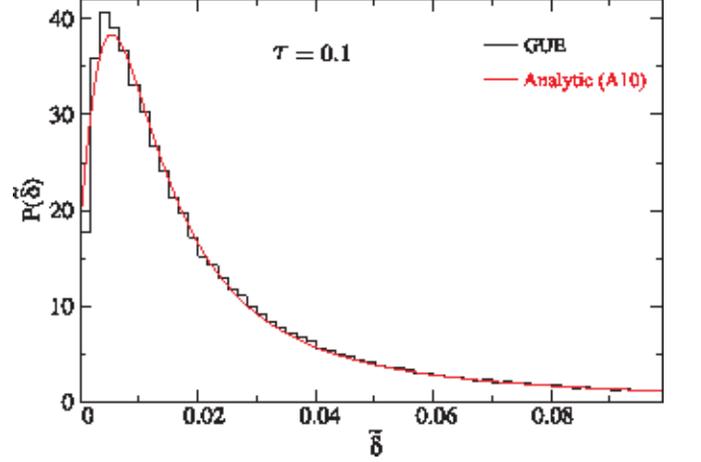}
        \caption{(Color online) Comparison of the distribution of the splittings of the eigenvalues of random matrices of the form Eq.~(\ref{eq:1}) with $H^{mixed}$ replaced by a random $300\times 300$ matrix from the GUE (black histogram) with the analytical result Eq.~(\ref{PDappr}) (full red line).}
\label{ptau}
\end{figure}

The joint probability density distribution of the eigenvalues of $2\times 2$ random matrices $H$ from the GUE reads
\begin{equation}
{\rm P}(e_1,e_2)=\frac{1}{\pi\tilde\sigma^2}(e_2-e_1)^2e^{-e_1^2/\tilde\sigma}e^{-e_2^2/\tilde\sigma}\, ,
\end{equation}
where $\tilde\sigma=2\sigma_H^2$ is given in terms of the variances of the matrix elements of $H$ (see below Eq.~(\ref{eq:mixedH})) and fixes the scale of the eigenvalues.
This yields for the probability, that one of its eigenvalues takes a value from the interval $[E_1\, E_1+{\rm d}E_1]$ 
\begin{eqnarray}
{\rm P}(E_1){\rm d}E_1&=&\int_{-\infty}^{\infty}{\rm d}e_1\int_{-\infty}^{\infty}{\rm d}e_2{\rm P}(e_1,e_2)\delta(E_1-e_1){\rm d}E_1\nonumber\\
&=&\frac{1}{2}\frac{1}{\sqrt{\pi\tilde\sigma}}e^{-E_1^2/\tilde\sigma}\left[1+2\frac{E_1^2}{\tilde\sigma}\right]dE_1\, .
\label{Pe1}
\end{eqnarray}
The limiting values of $E_1$ are given by the radius of the Wigner semicircle, $-R\leq E_1\leq R,\, R=2\sqrt{N}\sigma_H$ with $N$ the dimension of $H$, i.e., $N=2$.
Furthermore, the nearest-neighbor level-spacing distribution is obtained as 
\begin{eqnarray}
{\rm P}(s)&=&\int_{-\infty}^{\infty}{\rm d}e_1\int_{-\infty}^{\infty}{\rm d}e_2{\rm P}(e_1,e_2)\delta(s-\vert e_1-e_2\vert)\nonumber\\
&=&\frac{1}{\sqrt{2\pi\tilde\sigma}}\frac{2s^2}{\tilde\sigma}e^{-s^2/2\tilde\sigma}\, ,
\label{Ps}
\end{eqnarray}
yielding for the mean spacing
\begin{equation}
\bar s=4\sqrt{\frac{\tilde\sigma}{2\pi}}\, .
\end{equation}
The distribution of the quantities $\mathcal{W}_i$ defined above Eq.~(\ref{Eq:EV}) is given by
\begin{equation}
{\rm P}(\mathcal{W})=\frac{\mathcal{W}}{\sigma_{v2}^4}e^{-\mathcal{W}/\sigma_{v2}}.
\label{Pw}
\end{equation}
Here, $\sigma_{v2}$ is the variance of the matrix element $V$ defined in Eq.~(\ref{eq:1}). We may express it in terms of the variance of the matrix elements of $H$,
\begin{equation}
\sigma_{v2}=\tau\sigma_H\, .
\label{scale_sigma}
\end{equation}
To derive the distribution Eq.~(\ref{Pw}) we used the fact that the entries of $V$ are Gaussian distributed with zero mean. From the results Eq.~(\ref{Pe1}) and Eq.~(\ref{Pw}) we obtain for the probability distribution of the splittings Eq.~(\ref{DLF2A}) 
\begin{eqnarray}
&&{\rm P}(\Delta){\rm d}\Delta\\
&&=\int_{-R}^{R}{\rm de}{\rm P}(e)\int_0^\infty{\rm d}\mathcal{W}{\rm P}(\mathcal{W})\delta\left(\Delta-\frac{\mathcal{W}}{\vert e\vert}\right){\rm d}\Delta\nonumber\\
&&=\frac{1}{\sqrt{\pi\tilde\sigma}}\frac{\Delta}{\sigma_{v2}^4}\int_0^R{\rm d}ee^2e^{-e^2/\tilde\sigma}\left(1+2\frac{e^2}{\tilde\sigma}\right)e^{-\Delta e/\sigma_{v2}^2}{\rm d}\Delta.\nonumber
\label{PD}
\end{eqnarray}
Expressing $\tilde\sigma$ and $\sigma_{v2}$ in terms of $\sigma_H$, rescaling the splittings $\Delta\rightarrow\sigma_H\Delta$ and substituting the integration variable $e$ by $y=e\sigma_H$ leads to a scale-invariant expression for the splitting distribution, 
\begin{equation}
{\rm P}(\Delta)=\frac{\Delta}{\sqrt{2\pi}}\frac{1}{\tau^4}\int_0^{\infty}{\rm d}yy^2e^{-y^2/2}\left(1+y^2\right)e^{-\Delta y/\tau^2}
\label{PDrsk}
\end{equation}
where we replaced the upper integration limit $y\leq R/\sigma_H=2\sqrt{N}$ by $\infty$ using that the integrand becomes negligibly small for $y\gtrsim R/\sigma_H$. In Fig.~\ref{ev_verteilung} we compare this analytical result for $\tau=0.1$ with the distribution of $\mathcal{W}_l/\Delta_l$ obtained for an ensemble of $10000$ random matrices of the form Eq.~(\ref{eq:1}) with $H^{mixed}$ replaced by a random $2\times 2$ matrix from the GUE. The agreement is very good.
A simpler expression, which, however provides an approximation as good as Eq.~(\ref{PDrsk}) is obtained, when we replace in Eq.~(\ref{DLF2A}) the quantity $\mathcal{W}_l$ by its average $\bar{\mathcal{W}}=2\sigma_{v2}^2$. Here, we have to take into account that due to the limited range of values for $E_l$ the ratio $\bar{\mathcal{W}}/\vert E_l\vert\geq\bar{\mathcal{W}}/R$ is bordered from below, whereas $\Delta_l$ in Eq.~(\ref{DLF2A}) may take any value between $0$ and $\infty$. Then we obtain
\begin{eqnarray}
&&{\rm P}(\Delta){\rm d}\Delta\\
&&=\frac{\bar{\mathcal{W}}}{(\Delta+\frac{\bar{\mathcal{W}}}{R})^2}\frac{1}{\sqrt{\pi\tilde\sigma}}e^{-\frac{\bar{\mathcal{W}}^2}{\tilde\sigma(\Delta+\frac{\bar{\mathcal{W}}}{R})^2}}\left(1+\frac{2}{\tilde\sigma}\frac{\bar{\mathcal{W}}^2}{(\Delta+\frac{\bar{\mathcal{W}}}{R})^2}\right){\rm d}\Delta.\nonumber
\label{PD0}
\end{eqnarray}
Inserting the relation Eq.~(\ref{scale_sigma}) and again rescaling the splittings $\Delta\rightarrow\sigma_H\Delta$ finally leads to 
\begin{equation}
{\rm P}(\Delta)=\frac{2}{\sqrt{\pi}}\frac{\tau^2}{(\Delta+\tau^2)^2}e^{-4\frac{\tau^4}{(\Delta+\tau^2)^2}}\left[1+8\frac{\tau^4}{(\Delta+\tau^2)^2}\right].
\label{PDappr}
\end{equation}
In Fig.~\ref{ev_verteilung} we compare this result for $\tau=0.1$ with the numerically obtained distribution of the ratios $\mathcal{W}_l/E_l$ and the analytical expression Eq.~(\ref{PDrsk}) and in Fig.~\ref{ptau} we compare it with the splitting distribution obtained for an ensemble of $100$ random matrices of the form Eq.~(\ref{eq:1}) where we replaced $H^{mixed}$ by a $300\times 300$-dimensional random matrices from the GUE. The very good agreement demonstrates that the expression Eq.~(\ref{PDappr}) indeed provides a description for the distribution of the splittings deduced from the RMT model Eq.~(\ref{eq:1}), and thus of the eigenvalues of constant-width billiards. Note that like the Cauchy distribution the distributions Eq.(\ref{PDrsk}) and (\ref{PDappr}) are normalizable but possess no finite moments.     

\end{document}